\documentclass[aps,prl,preprintnumbers,twocolumn,groupedaddress,nofootinbib]{revtex4}

\usepackage[dvips]{graphicx}
\usepackage{color}
\usepackage{amsmath,amssymb,slashed}
\usepackage{tikz}
\usepackage{hyperref}

\newif\ifdraft
\drafttrue
\newif\ifpreprint
\preprinttrue
\def\spa#1.#2{\left\langle#1\,#2\right\rangle}
\def\spb#1.#2{\left[#1\,#2\right]}

\def\beq{\begin{equation}}
\def\eeq{\end{equation}}
\let\Re\relax
\let\Im\relax
\DeclareMathOperator{\Re}{Re}
\DeclareMathOperator{\Im}{Im}

\newcommand{\eq}{\begin{equation}}
\newcommand{\eqe}{\end{equation}}
\newcommand{\eqa}{\begin{eqnarray}}
\newcommand{\eqae}{\end{eqnarray}}

\newcommand{\bea}{\begin{eqnarray}}
\newcommand{\eea}{\end{eqnarray}}
\newcommand{\dd}{\mathrm{d}}

\newcommand{\RR}{\mathbb R}

\newcommand{\NN}{\mathbb N}

\newbox\charbox
\newbox\slabox
\def\s#1{{      % Feynman slash
        \setbox\charbox=\hbox{$#1$}
        \setbox\slabox=\hbox{$/$}
        \dimen\charbox=\ht\slabox
        \advance\dimen\charbox by -\dp\slabox
        \advance\dimen\charbox by -\ht\charbox
        \advance\dimen\charbox by \dp\charbox
        \divide\dimen\charbox by 2
        \raise-\dimen\charbox\hbox to \wd\charbox{\hss/\hss}
        \llap{$#1$}
}}

\begin{document}

\preprint{UUITP--34/19}

\title{
All-order $\alpha'$-expansion of one-loop open-string integrals
}

\author{Carlos R.\ Mafra$^a$ and
Oliver Schlotterer$^{b}$}
\affiliation{$^a$ STAG Research Centre and Mathematical Sciences, University of Southampton,
Highfield, Southampton SO17 1BJ, UK
}
\affiliation{$^b$ Department of Physics and Astronomy, Uppsala University, 75108 Uppsala, Sweden.}

\begin{abstract}
We present a new method to evaluate the $\alpha'$-expansion
of genus-one integrals over open-string punctures and unravel the structure
of the elliptic multiple zeta values in its coefficients. This is done by obtaining
a simple differential equation of Knizhnik--Zamolodchikov--Bernard-type satisfied by
generating functions of such integrals, and solving it via Picard iteration.
The initial condition involves the generating functions at the cusp $\tau\to
i\infty$ and can be reduced to genus-zero integrals.
\end{abstract}

\maketitle

%%%%%%%%%%%%%%%%%%%%%%%%%%%%%%%%%%%%%%%%%%%%%%%%%%%%%%%%%%%%
\section{Introduction}
%%%%%%%%%%%%%%%%%%%%%%%%%%%%%%%%%%%%%%%%%%%%%%%%%%%%%%%%%%%%

\vspace{-0.3cm}
\noindent
Elliptic analogues of polylogarithms \cite{Lev, BrownLevin} and multiple zeta
values \cite{Enriquez:Emzv} have become a driving force in
higher-order computations of scattering amplitudes in quantum field theories
and string theories. The study of their differential equations and their connections with
modular forms turned into a vibrant research area at the interface of particle phenomenology,
string theory and number theory. In the same way as a variety of Feynman integrals has
been recently expressed in terms of elliptic polylogarithms and iterated integrals of modular
forms \cite{Feynrefs, Adams:2018yfj},
the low-energy expansion of one-loop open-string amplitudes introduces elliptic multiple zeta values
(eMZVs) \cite{Broedel:2014vla, Broedel:2017jdo, Broedel:2019vjc}.

So far, the appearance of eMZVs in one-loop open-string amplitudes arose from
direct integration
over the punctures on a genus-one worldsheet of cylinder or
M\"obius-strip topology.
Although there is no conceptual bottleneck in
extending the techniques of \cite{Broedel:2014vla, Broedel:2017jdo, Broedel:2019vjc}
to arbitrary multiplicities and orders in the inverse
string tension $\alpha'$,
in this letter we will present
a new method to evaluate these genus-one integrals.

% NEW VERSION
Our method automatically generates the eMZVs in their minimal form \cite{Enriquez:Emzv, Broedel:2015hia} 
and reveals elegant structures in the $\alpha'$-expansions. It rests on new differential equations for 
genus-one integrals which open up connections with modern number-theoretic concepts:
elliptic associators \cite{KZB} -- generating series
of eMZVs -- and Tsunogai's derivations dual to Eisenstein series \cite{Tsunogai}
which govern the counting of independent eMZVs \cite{Broedel:2015hia}. More details will be given in a longer
companion paper \cite{bigpaper}.

\begin{figure}[b]
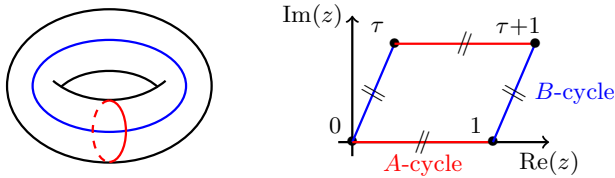

\begin{center}
\tikzpicture[scale=0.34, line width=0.30mm]
\draw(0,0) ellipse  (4cm and 3cm);
\draw(-2.2,0.2) .. controls (-1,-0.8) and (1,-0.8) .. (2.2,0.2);
\draw(-1.9,-0.05) .. controls (-1,0.8) and (1,0.8) .. (1.9,-0.05);
\draw[blue](0,0) ellipse  (3cm and 1.8cm);
\draw[red] (0,-2.975) arc (-90:90:0.65cm and 1.2cm);
\draw[red,dashed] (0,-0.575) arc (90:270:0.65cm and 1.2cm);
\scope[xshift=9.5cm,yshift=-2.2cm,scale=1.1]
\draw[->](-0.5,0) -- (7,0) node[below]{${\rm Re}(z)$};
\draw[->](0,-0.5) -- (0,4.5) node[left]{${\rm Im}(z)$};
\draw(0,0)node{$\bullet$};
\draw(-0.6,0.6)node{$0$};
\draw[blue](0,0) -- (1.5,3.5);
\draw(0.75,1.75)node[rotate=60]{$| \; \! \!|$};
\draw (1.5,3.5)node{$\bullet$} ;
\draw(0.9,4.1)node{$\tau$};
\draw[red](0,0) -- (5,0);
\draw(2.5,0)node[rotate=-20]{$| \; \! \! |$};
\draw (5,0)node{$\bullet$};
\draw (4.4,0.6)node{$1$};
\draw[red](1.5,3.5) -- (6.5,3.5);
\draw(4,3.5)node[rotate=-20]{$| \; \! \! |$};
\draw[blue](5,0) -- (6.5,3.5);
\draw(5.75,1.75)node[rotate=60]{$| \; \! \! |$};
\draw(6.5,3.5)node{$\bullet$};
\draw(5.9,4.1)node{$\tau{+}1$};
\draw[red](2.5,-0.8)node{$A$-cycle};
\draw[blue](7.9,1.75)node{$B$-cycle};
\endscope
\endtikzpicture
\caption{We parameterize the torus through the lattice $\frac{\Bbb C}{\Bbb Z {+} \tau \Bbb Z}$
with identifications $z \cong z{+}1 \cong z{+}\tau$ along the $A$- and $B$-cycle.}
\label{figtorus}
\end{center}
\end{figure}

\vspace{-0.15cm}

%%%%%%%%%%%%%%%%%%%%%%%%%%%%%%%%%%%%%%%%%
\section{Open-string integrals at genus one}
%%%%%%%%%%%%%%%%%%%%%%%%%%%%%%%%%%%%%%%%%

\vspace{-0.3cm}
\noindent
One-loop string amplitudes are described by correlation
functions of vertex operators in a conformal field theory over a genus-one
Riemann surface, the torus. The location of the vertex operator associated
with the $j^{\rm th}$ external string state is parameterized by
the coordinates $z_j=u_j\tau{+}v_j$ with $u_j ,v_j \in (0,1)$,
where $\tau$ is the modulus with $\Im \tau>0$, see figure \ref{figtorus}, and we define $z_{ij}\equiv z_i-z_j$.
 
\begin{figure}[h]
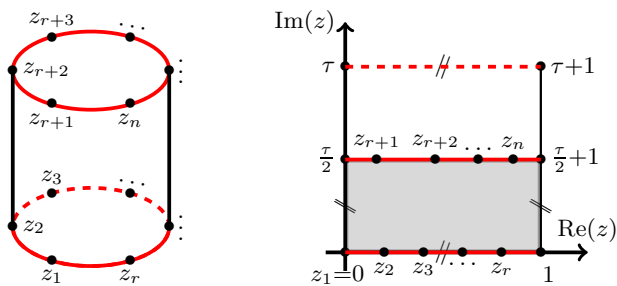

\begin{center}
\tikzpicture [scale=0.52,line width=0.50mm]
\draw[red] (2,-4) ellipse (2cm and 1cm);
\draw[white,fill=white] (0,-2.8) rectangle (4,-3.9);
\draw[red,dashed] (2,-4) ellipse (2cm and 1cm);
\draw[red] (2,0) ellipse (2cm and 1cm);
\draw (0,0) -- (0,-4);
\draw (4,0) -- (4,-4);
%%%%%%
\draw (1,0.85)node{$\bullet$}node[above]{$z_{r{+}3}$};
\draw (3,0.85)node{$\bullet$}node[above,rotate=-10]{$\ldots$};
\draw (4,0) node{$\bullet$};
\draw (4.3,0.18) node{$\vdots$};
\draw (0,0) node{$\bullet$}node[right]{$z_{r{+}2}$};
\draw (1,-0.85) node{$\bullet$}  node[below]{$z_{r{+}1}$};
\draw (3,-0.85) node{$\bullet$}  node[below]{$z_n$};
\scope[yshift=-4cm]
\draw (1,0.85)node{$\bullet$}node[above]{$z_3$};
\draw (3,0.85)node{$\bullet$}node[above,rotate=-10]{$\ldots$};
\draw (4,0) node{$\bullet$} ;
\draw (4.3,0.18) node{$\vdots$};
\draw (0,0) node{$\bullet$}node[right]{$z_2$};
\draw (1,-0.85) node{$\bullet$}  node[below]{$z_1$};
\draw (3,-0.85) node{$\bullet$}  node[below]{$z_r$};
\endscope
%%%%%%%%%%%%%%%
%%%%%%%%%%%%%%%
\scope[xshift=8.5cm,yshift=-4.65cm,yscale=0.95]
\draw[->](-0.5,0) -- (6.2,0) node[above]{${\rm Re}(z)$};
\draw[->](0,-0.5) -- (0,6.2) node[left]{${\rm Im}(z)$};
\draw[line width=0.3mm](5,0) -- (5,5) ;
\draw(0,0)node{$\bullet$};
\draw(-0.2,-0.6)node{$z_1{=}0$};
\draw(0,1.25)node[rotate=60]{$| \! |$};
\draw(5,1.25)node[rotate=60]{$| \! |$};
\draw(2.5,0)node[rotate=-20]{$| \; \! \! |$};
\draw(2.5,5)node[rotate=-20]{$| \; \! \! |$};
\draw (0,5)node{$\bullet$}node[left]{$\tau$};
\draw (0,2.5)node{$\bullet$}node[left]{$\tfrac{\tau}{2}$};
\draw (5,2.5)node{$\bullet$}node[right]{$\tfrac{\tau}{2}{+}1$};
\draw (5,5)node{$\bullet$}node[right]{$\tau{+}1$};
\draw[fill=gray, opacity=0.3,line width=0.3mm] (0.05,0.05) rectangle (4.95,2.45);
\draw[red](0,0) -- (5,0);
\draw[red](0,2.5) -- (5,2.5);
\draw[red,dashed](0,5) -- (5,5);
\draw (5,0)node{$\bullet$};
\draw (5.2,-0.6)node{$1$};
\draw(1,0)node{$\bullet$}node[below]{$z_2$};
\draw(2,0)node{$\bullet$}node[below]{$z_3$};
\draw(3,0)node{$\bullet$}node[below]{$\ldots$};
\draw(4,0)node{$\bullet$}node[below]{$z_r$};
\draw(0.8,2.5)node{$\bullet$}node[above]{$z_{r+1}$};
\draw(2.3,2.5)node{$\bullet$}node[above]{$z_{r+2}$};
\draw(3.4,2.5)node{$\bullet$}node[above]{$\ldots$};
\draw(4.3,2.5)node{$\bullet$}node[above]{$z_n$};
\endscope
%%%%%%%%%%%%%%
%%%%%%%%%%%%%%
\endtikzpicture
\caption{The cylinder parameterization.}
\label{fig1}
\end{center}
\end{figure}

By suitable involutions of the torus \cite{polchinski1},
one obtains the surfaces describing the scattering of open-string states, the
cylinder and the M\"obius strip.
The two boundaries of the cylinder will be parameterized by
the $A$-cycle $z_j \in (0,1)$ and its displacement $z_j \in \tfrac{\tau}{2}{+}(0,1)$ by
half a $B$-cycle, i.e.\ $u_j {\in} \{0,\frac{1}{2}\}$ and $\dd z_j {=} \dd v_j$. See figure \ref{fig1}.

\vspace{-0.2cm}

\vspace{-0.48cm}

\medskip
The massless $n$-point one-loop amplitudes
of the open superstring give rise to integrals of the form ($z_1=0$)
\cite{Broedel:2014vla}
\beq
\label{needInt}
\hskip-4pt\int_{{\cal C}(\ast)}\!\!  \Big( \prod_{j=2}^n \dd z_j \Big) \,
f_{i_1j_1}^{(k_1)} f_{i_2 j_2}^{(k_2)}\cdots
\exp \Big( \sum_{i<j}^n s_{ij} {\cal G}(z_{ij},\tau) \Big),
\eeq
with differing integration domains ${\cal C}(\ast)$ for the cylinder and the M\"obius
strips. For planar cylinders, we set $\ast \rightarrow 1,2,\ldots,n$ and parametrize the
domain as
\beq
{\cal C}(1,2,\ldots,n) = \{ z_{j=2,\ldots,n} \in \RR, \
0{<}z_2{<}\ldots{<}z_n{<}1\}\,,
\eeq
see figure \ref{fig1} and \cite{bigpaper} for the non-planar analogue
with $\ast \rightarrow \begin{smallmatrix} r{+}1,\ldots,n \\ 1,2,3,\ldots,r \end{smallmatrix}$.
Furthermore, in the integrand of \eqref{needInt},
$f^{(k)}_{ij}\equiv f^{(k)}(z_{ij},\tau)$ denote the
Laurent coefficients of the doubly-periodic Kronecker--Eisenstein series
defined by \cite{Kroneck, BrownLevin}
\begin{align}
\Omega(z,\eta,\tau) &= \exp \Big( 2\pi i \eta\frac{ \Im z }{\Im \tau} \Big)
\frac{ \theta_1'(0,\tau) \theta_1(z+\eta,\tau) }{\theta_1(z,\tau)
\theta_1(\eta,\tau)}\,,\\
\label{let10}
\Omega(z,\eta,\tau) &=  \sum_{k=0}^{\infty} \eta^{k-1} f^{(k)}(z,\tau) \, ,
\end{align}
%% NEW INFO
which generates the integration kernels of multiple elliptic polylogarithms \cite{BrownLevin}.
The simplest examples of the coefficient
functions are $f^{(0)}(z,\tau)=1$ and
$f^{(1)}(z,\tau)=\partial_z \log \theta_1(z,\tau)+2\pi i \frac{ \Im z }{\Im \tau}$, and higher
$f^{(k\geq 2)}(z,\tau)$ do not have any poles in $z$.

Finally, $\exp \big( \sum_{i<j}^n s_{ij} {\cal G}(z_{ij},\tau) \big)$ in \eqref{needInt} is the
Koba--Nielsen factor written in terms of dimensionless Mandelstam invariants
$s_{ij} = - 2\alpha' k_i \cdot k_j$
and Green functions ${\cal G}(z,\tau)$
subject to the universal differential equation
\begin{align}
\partial_{v_i} {\cal G}(z_{ij},\tau) &= - f^{(1)}(z_{ij},\tau) \label{let13} \\
%%%
2\pi i \partial_\tau {\cal G}(z_{ij},\tau) &= - f^{(2)}(z_{ij},\tau) - 2 \zeta_2\,, \notag
\end{align}
where $\partial_{v_i}$ is the derivative along the cylinder boundary,
and $\zeta_n {=} \sum_{k=1}^{\infty} \frac{1}{k^{n}}$ with $n{\geq} 2$ denote Riemann
zeta values.

\medskip
{\bf A. Generating functions:}
Instead of handling the $\alpha'$-expansion of the individual integrals \eqref{needInt} as in the
method of \cite{Broedel:2014vla, Broedel:2017jdo, Broedel:2019vjc},
we will evaluate the following generating function of integrals (with $\eta_{23 \ldots n} = \eta_{2}{+}
\eta_{3}{+}\ldots{+}\eta_{n}$)
\begin{align}
&Z^\tau_{\vec{\eta}}( \ast | 1,2,\ldots,n) = \! \int \limits_{{\cal C}(\ast)} \prod_{j=2}^n \dd z_j
\exp \Big( \sum_{i<j}^n s_{ij} {\cal G}(z_{ij},\tau) \Big) \! \!
\label{let7}\\
%%%
&\ \ \times
 \Omega( z_{12},\eta_{23 \ldots n} ,\tau)
  \Omega( z_{23},\eta_{3 \ldots n} ,\tau)
  \ldots
   \Omega( z_{n-1,n},\eta_{ n}  ,\tau) \, . \notag
\end{align}
The integrands $f_{i_1j_1}^{(k_1)} f_{i_2 j_2}^{(k_2)}\cdots $ in \eqref{needInt} relevant to $n$-point open-superstring
amplitudes have $k_1{+}k_2{+}\ldots = n{-}4$ and reside at the order of $\eta_j^{-3}$ of \eqref{let7}.
Moreover, $(n{\geq }8)$-point integrands additionally involve holomorphic
Eisenstein series ${\rm G}_{\ell \geq 4}(\tau)= {-} f^{(\ell)}(0,\tau)$ \cite{Broedel:2014vla}
multiplying \eqref{needInt} at $k_1{+}k_2{+}\ldots = n{-}4{-}\ell$ as seen
at the $\eta_j^{-3-\ell}$-order of~\eqref{let7}.

Although the cylinder contribution to one-loop open-string amplitudes is
localized at purely imaginary $\tau$ as drawn in figure \ref{fig1}, we will define
and evaluate the integrals (\ref{let7}) for generic $\tau$ in the upper half
plane with $\Re \tau \neq 0$. In view of the parental torus,
$Z^\tau_{\vec{\eta}}( 1,2,\ldots,n| \cdot)$ and $Z^\tau_{\vec{\eta}}(
\begin{smallmatrix} r{+}1,\ldots,n \\ 1,2,3,\ldots,r \end{smallmatrix}|
\cdot)$ will be referred to as planar and non-planar {\it $A$-cycle
integrals}, respectively.

M\"obius-strip integrals can be reconstructed by specializing planar $A$-cycle integrals to $\Re
 \tau=\frac{1}{2}$, and the cancellation of tadpole divergences from one-loop
 open-superstring amplitudes can be analyzed as in \cite{Green:1984ed}.

The $A$-cycle integrand \eqref{let7} at $n$ points involves
$n{-}1$ factors of the Kronecker--Eisenstein series (\ref{let10}) at different arguments.
The second entry $Z^\tau_{\vec{\eta}}( \ast |A)$ specifies permutations
$A=a_1 a_2\ldots a_n \in S_n$ of the arguments, and $\Omega(\ldots)$ at different $z_{a_j},\eta_{a_j}$
are related by the Fay identity~\cite{Fayref}
\begin{align}
\Omega(z_1,\eta_1,\tau)&\Omega(z_2,\eta_2,\tau) =
\Omega(z_1,\eta_1{+}\eta_2,\tau) \Omega(z_2{-}z_1,\eta_2,\tau) \notag \\
& \ \ \ \ \ \ \ +\Omega(z_2,\eta_1{+}\eta_2,\tau) \Omega(z_1{-}z_2,\eta_1,\tau)\,,
\label{let14}
\end{align}
% NEW STUFF
which can be thought of as a doubly-periodic generalization of the partial-fraction relation
$\frac{1}{z_1 z_2} {=} \frac{1}{(z_1{-}z_2)}( \frac{1}{z_2} {-}  \frac{1}{z_1} )$.
Repeated use of (\ref{let14}) and imposing $\eta_1= - \sum_{j=2}^n \eta_j$ only leaves
$(n{-}1)!$ independent permutations of the integrand in (\ref{let7}), and we will use a basis
of $Z^\tau_{\vec{\eta}}( \ast |1,B)$ with permutations $B \in S_{n-1}$ acting on $2,3,\ldots,n$.

\medskip
{\bf B. The differential equation:}
As will be derived in \cite{bigpaper}, the $\tau$-derivatives of (\ref{let7}) can be written as
\beq
2\pi i \partial_\tau Z^\tau_{\vec{\eta}}( A |1,B)
= \!\!\sum_{C \in S_{n-1}}  \! \! D^\tau_{\vec{\eta}}(B|C) Z^\tau_{\vec{\eta}}(A | 1,C) \, ,
\label{let15}
\eeq
where the $(n{-}1)!\times (n{-}1)!$ matrix $D^\tau_{\vec{\eta}}$ is a differential operator
w.r.t.\ $\eta_j$.
Its detailed form will be exemplified in the next section and
follows from the properties (\ref{let13}) of the Green function,
the vanishing of boundary terms $\int \dd v_j \, \partial_{v_j}(\ldots)$
and the mixed heat equation ($u,v \in \RR$)
\beq
2\pi i \partial_{\tau} \Omega(u\tau{+} v,\eta,\tau) = \partial_v \partial_\eta \Omega(u\tau{+} v,\eta,\tau)
\, .
\label{let16}
\eeq
Most importantly, the form of $D^\tau_{\vec{\eta}}(B|C)$ does not depend
on the planar or non-planar integration cycle $A$, and its entries are linear in
the dimensionless Mandelstam invariants $s_{ij}$ and therefore in $\alpha'$.

Hence, the $\alpha'$-expansion of the $A$-cycle integrals $Z^\tau_{\vec{\eta}}$ follows
from the solution of (\ref{let15}) via Picard iteration --
an infinite iteration of the integrated version
$Z^{\tau}_{\vec{\eta}}(A|1,B) = Z^{i\infty}_{\vec{\eta}}(A|1,B) + \int^\tau_{i\infty} \frac{ \dd \tau' }{2\pi i}
\sum_{C\in S_{n-1}} D_{\vec{\eta}}^{\tau'}(B|C)Z^{\tau'}_{\vec{\eta}}(A|1,C)$
of (\ref{let15}) such that
\begin{align}
 &Z^{\tau}_{\vec{\eta}}(A|1,B) = \sum_{k=0}^{\infty} \Big( \frac{1}{2\pi i } \Big)^k \int^{\tau}_{i\infty} \dd \tau_1   \int^{\tau_1}_{i\infty} \dd \tau_2 \ldots  \int^{\tau_{k-1}}_{i\infty} \dd \tau_k
\notag
\\
& \ \ \
\times \! \! \! \sum_{C  \in S_{n-1} } \! \! (D^{\tau_k}_{\vec{\eta}}\cdot \ldots \cdot D^{\tau_2}_{\vec{\eta}}\cdot D^{\tau_1}_{\vec{\eta}})(B|C)\,
Z^{i\infty}_{\vec{\eta}}(A|1,C)
\label{let2}
\end{align}
with matrix products $D^{\tau_k}_{\vec{\eta}}\cdot \ldots \cdot D^{\tau_2}_{\vec{\eta}}  D^{\tau_1}_{\vec{\eta}}$.
As an initial value, the degeneration
$Z^{i\infty}_{\vec{\eta}}$ at the cusp $\tau {\rightarrow} i \infty$ will be expressed
in terms of disk integrals with two additional punctures
from the pinching of the $A$-cycle in figure \ref{figtorus}.

As will be detailed in \cite{bigpaper}, the entire $\tau$-dependence
of $D^{\tau}_{\vec{\eta}}$ is carried by Weierstrass functions (with ${\rm
G}_0=-1$)
\beq
\wp(\eta,\tau) = -\frac{ {\rm G}_0 }{\eta^2} + \sum_{k=4}^\infty  (k{-}1) \eta^{k-2} {\rm G}_k(\tau)\,.
\label{let3}
\eeq
This allows us to decompose
\beq
D^{\tau}_{\vec{\eta}} =  \sum_{k=0}^\infty  (1{-}k)  {\rm G}_k(\tau)
r_{\vec{\eta}}(\epsilon_k) \,,
\label{let4}
\eeq
where $r_{\vec{\eta}}(\epsilon_k)$ are $(n{-}1)! {\times} (n{-}1)!$ matrices whose
entries are independent of $\tau$, rational functions of $\eta_j$, linear in $s_{ij}$
and may involve second derivatives $\partial_{\eta_i}\partial_{\eta_j}$.
Note that $r_{\vec{\eta}}(\epsilon_2)= 0$ and $r_{\vec{\eta}}(\epsilon_{2p-1}) = 0 \ \forall \
p \in \NN$ by (\ref{let3}).

\medskip
{\bf C. The main result:}
Based on (\ref{let4}), the open-string integrals (\ref{let2}) can be expressed in terms of
iterated Eisenstein integrals
\beq
\gamma(k_1,k_2,\ldots,k_r|\tau) =  \int^{i \infty}_{\tau} \frac{ \dd \tau' }{2\pi i } \,
 {\rm G}_{k_r}(\tau' )\, \gamma(k_1,\ldots,k_{r-1}|\tau')
   \label{let5}
\eeq
subject to $\gamma(\emptyset|\tau)=1$ and tangential-base-point regularization \cite{Brown:mmv},
e.g.\ $\gamma(0|\tau)= \frac{ \tau}{2\pi i}$. As the main result of this work, we can therefore
bring the open-string $\alpha'$-expansion into the following elegant form:
\begin{align}
Z^{\tau}_{\vec{\eta}} &(A|1,B)= \sum_{r=0}^{\infty}
 \sum_{k_1,k_2,\ldots,k_r \atop{=0,4,6,8,\ldots} }
\!  \gamma(k_1,k_2,\ldots,k_r|\tau)   \label{let6}\\
&\! \! \!   \times \,  \prod_{j=1}^r (k_j{-}1)\! \! \sum_{C\in S_{n-1}} \!
r_{\vec{\eta}}(\epsilon_{k_r} \ldots  \epsilon_{k_2}  \epsilon_{k_1})_B{}^C
Z^{i\infty}_{\vec{\eta}}(A|1,C)\,,
\notag
\end{align}
where $r_{\vec{\eta}}( \epsilon_{k_r} \ldots  \epsilon_{k_2}\epsilon_{k_1}) {\equiv}
r_{\vec{\eta}}(\epsilon_{k_r}) \cdots
r_{\vec{\eta}}(\epsilon_{k_2})r_{\vec{\eta}}(\epsilon_{k_1})$.
% OLD VERSION
%Since each order in $\alpha'$ is expressible in terms of eMZVs
%\cite{Broedel:2014vla, Broedel:2017jdo, Broedel:2019vjc}, the
%$r_{\vec{\eta}}(\epsilon_k)$ should be matrix representations of Tsunogai's
%derivations $\epsilon_{k}$ dual to Eisenstein series \cite{Tsunogai}.
%
%
%%% NEW VERSION
Since each order in $\alpha'$ is expressible in terms of eMZVs
\cite{Broedel:2014vla, Broedel:2017jdo, Broedel:2019vjc} but not all of the
$\gamma(\ldots)$ are constructible from eMZVs, the
$r_{\vec{\eta}}(\epsilon_k)$ must obey certain commutation relations.
More specifically, the $r_{\vec{\eta}}(\epsilon_k)$ should preserve
the commutation relations of Tsunogai's derivations $\epsilon_{k}$ dual to 
Eisenstein series \cite{Tsunogai} which select the $\gamma(\ldots)$
with a realization via eMZVs \cite{Broedel:2015hia}. Hence, the $r_{\vec{\eta}}(\epsilon_k)$
are believed to furnish matrix representations of Tsunogai's derivations.
%%% END OF NEW VERSION
In particular, (\ref{let4}) brings the differential equation \eqref{let15} of
$Z^{\tau}_{\vec{\eta}}$ into the same form as that of the elliptic
Knizhnik--Zamolodchikov--Bernard associator
% NEW 
whose $\tau$-derivative involves the derivations $\epsilon_{k}$ acting on 
its non-commutative arguments \cite{KZB}.
% OLD
%where the derivations $\epsilon_{k}$ act on its non-commutative arguments.

The decomposition of eMZVs into iterated Eisenstein integrals automatically
incorporates all their relations over the rational numbers \cite{Broedel:2015hia}.
Moreover, the derivation of (\ref{let6}) does not rely on
any relation among the Mandelstam invariants. The $n$-point results of this work are valid
for $\frac{1}{2}n(n{-}1)$ independent $s_{ij}$, and one can still impose
momentum conservation when applying the $\alpha'$-expansion of $Z^{\tau}_{\vec{\eta}}$ to
string amplitudes.

\vspace{-0.15cm}

%%%%%%%%%%%%%%%%%%%%%%%%%%%%%%%%%%%%%%%%%
\section{Examples for differential operators}
%%%%%%%%%%%%%%%%%%%%%%%%%%%%%%%%%%%%%%%%%

\vspace{-0.3cm}
\noindent
%In this section, we present $(n{<} 4)$-point examples of the matrix-valued differential
%operators $D^\tau_{\vec{\eta}}$ in (\ref{let15}), and the four-point case is relegated
%to the appendix. All-multiplicity expressions as well as detailed derivations of the differential
%equations can be found in \cite{bigpaper}.
In this section, we present $(n{\leq} 3)$-point examples of the matrix-valued differential
operators $D^\tau_{\vec{\eta}}$ in (\ref{let15}). All-multiplicity expressions as well as 
detailed derivations of the differential equations can be found in \cite{bigpaper} (see
e.g.\ section 4.2 in the reference for the four-point case).

\medskip
{\bf A. Two points} allow for a single planar and non-planar $A$-cycle integral (\ref{let7}) each,
\begin{align}
Z^\tau_{\eta_2}(1,2|1,2) &= \int_0^1 \dd v_2 \,\Omega(v_{12},\eta_2,\tau) \,e^{s_{12}{\cal G}(v_{12},\tau)} \label{let19}\\
Z^\tau_{\eta_2}\big( \begin{smallmatrix} 2 \\ 1 \end{smallmatrix}|1,2\big) &= \int_0^1 \dd v_2 \,\Omega(v_{12}{+}\tfrac{\tau}{2},\eta_2,\tau) \,e^{s_{12}{\cal G}(v_{12}{+}\tfrac{\tau}{2},\tau)} \, .
\notag
\end{align}
Their $\tau$-derivatives resulting from (\ref{let13}), (\ref{let16}) and integration by parts w.r.t.\ $v_2$
take the universal form
\begin{align}
2\pi i \partial_\tau Z^\tau_{\eta_2}(\ast|1,2)&= s_{12}  \Big( \frac{1}{2} \partial_{\eta_2}^2 {-} \wp(\eta_2,\tau)  {-} 2 \zeta_2 \Big) Z^\tau_{\eta_2}(\ast|1,2)  \, ,  \label{let20}
\end{align}
so one can read off the scalar differential operator in (\ref{let15})
and the resulting representation of the derivations,
\begin{align}
\label{let21}
&D^\tau_{\eta_2}(2|2)=s_{12}  \Big( \frac{1}{2} \partial_{\eta_2}^2 -
\wp(\eta_2,\tau)  - 2 \zeta_2 \Big) \,, \\
& r_{\eta_2}(\epsilon_{0})  = s_{12} \Big( \frac{ 1}{\eta_2^{2}} {+}
 2 \zeta_2 {-} \frac{1}{2} \partial_{\eta_2}^2\Big)\,,\quad
r_{\eta_2}(\epsilon_{k\geq 4})  = s_{12} \eta_2^{k-2} \, .\notag
\end{align}
Note that various combinations of iterated Eisenstein integrals drop out from
the two-point instance of (\ref{let6}) since commutators
$[r_{\eta_2}(\epsilon_{k_1}) , r_{\eta_2}(\epsilon_{k_2}) ]$ with $k_1,k_2\geq
4$ vanish.

%%%%%%%%%%%%%%%%%%%%%%%%%%%%%%%%%%%%%%%%%%%%
%%%%%%%%%%%%%%%%%%%%%%%%%%%%%%%%%%%%%%%%%%%%
\medskip
{\bf B. Three points} give rise to $A$-cycle integrals
\begin{align}
&Z_{\eta_2,\eta_3}^{\tau}(\ast |1,2,3) = \int_{{\cal C}(\ast)} \dd z_2\, \dd z_3 \, \Omega(z_{12},\eta_2{+}\eta_3,\tau)
\label{let22}
\\
& \ \ \times
\Omega(z_{23},\eta_3,\tau)  e^{s_{12}{\cal G}(z_{12},\tau)+s_{13}{\cal G}(z_{13},\tau)+s_{23}{\cal G}(z_{23},\tau)}
\notag
\end{align}
that mix under $\tau$-derivatives ($s_{12\ldots p}\equiv \sum_{1\leq i<j}^p s_{ij}$),
\begin{align}
&2\pi i \partial_\tau Z_{\eta_2,\eta_3}^{\tau}(\ast |1,2,3) =
\Big(
{-}2 \zeta_2 s_{123}
\label{let23}\\
 &\ \ \ \ \ \ +  s_{12} \big[ \tfrac{1}{2} \partial^2_{\eta_2}- \wp(\eta_2{+}\eta_3,\tau) \big]  + s_{13} \big[ \tfrac{1}{2} \partial^2_{\eta_3}- \wp(\eta_3,\tau) \big]  \notag \\
&\ \ \ \ \ \  + s_{23}  \big[ \tfrac{1}{2} (\partial_{\eta_2}{-}\partial_{\eta_3})^2- \wp(\eta_3,\tau) \big]   \Big)  Z_{\eta_2,\eta_3}^{\tau}(\ast |1,2,3)   \notag \\
&\ \ + s_{13} \big[ \wp(\eta_2{+}\eta_3,\tau) - \wp(\eta_3,\tau) \big]
Z_{\eta_2,\eta_3}^{\tau}(\ast|1,3,2) \, .\notag
\end{align}
The resulting matrix entries of the $2\times 2$ differential operator in (\ref{let15}) read
\begin{align}
&D^{\tau}_{\eta_2,\eta_3}(2,3|2,3) = {-}2 \zeta_2 s_{123}
 + s_{12} \big[ \tfrac{1}{2} \partial^2_{\eta_2}{-} \wp(\eta_2{+}\eta_3,\tau) \big]
\notag \\
& \! \! + s_{23}  \big[ \tfrac{1}{2} (\partial_{\eta_2}{-}\partial_{\eta_3})^2{-} \wp(\eta_3,\tau) \big]
 + s_{13} \big[ \tfrac{1}{2} \partial^2_{\eta_3}{-} \wp(\eta_3,\tau) \big] \notag \\
&D^{\tau}_{\eta_2,\eta_3}(2,3|3,2) =
s_{13} \big[ \wp(\eta_2{+}\eta_3,\tau) {-} \wp(\eta_3,\tau) \big]  \, , \label{let24}
\end{align}
and the first row is always sufficient to generate the remaining entries via permutations
of $s_{ij}$ and $\eta_j$, e.g.
\begin{align}
D^{\tau}_{\eta_2,\eta_3}(3,2|3,2)&=D^{\tau}_{\eta_2,\eta_3}(2,3|2,3)   \, \big|^{s_{12}\leftrightarrow s_{13}}_{\eta_2 \leftrightarrow \eta_3}
  \label{let25} \\
D^{\tau}_{\eta_2,\eta_3}(3,2|2,3)&=D^{\tau}_{\eta_2,\eta_3}(2,3|3,2)   \, \big|^{s_{12}\leftrightarrow s_{13}}_{\eta_2 \leftrightarrow \eta_3}\, . \notag
\end{align}
One can read off the $2\times 2$ matrix representations of the derivations ($k\neq 2$),
\begin{align}
r_{\eta_2,\eta_3}(\epsilon_k) &= \delta_{k,0}\Big( 2 \zeta_2 s_{123}
- \frac{1}{2} s_{23} (\partial_{\eta_2} {-} \partial_{\eta_3})^2
- \frac{1}{2} s_{12} \partial_{\eta_2}^2  \notag \\
& \ \ \ \ \ \  - \frac{1}{2} s_{13} \partial_{\eta_3}^2  \Big) 1_{2\times 2}
+ \eta^{k-2}_{23} \Big( \begin{smallmatrix} s_{12} &-s_{13} \\-s_{12}&s_{13}  \end{smallmatrix} \Big) \label{let25A} \\
& \ \ \ 
+ \eta^{k-2}_{2} \Big( \begin{smallmatrix} 0 &0 \\ s_{12} &s_{12}{+}s_{23}  \end{smallmatrix} \Big)
+ \eta^{k-2}_{3} \Big( \begin{smallmatrix} s_{13}{+}s_{23} & s_{13} \\0 &0  \end{smallmatrix} \Big)
\, , \notag
\end{align}
where $[ r_{\eta_2,\eta_3}(\epsilon_{k_1\geq 4}),
r_{\eta_2,\eta_3}(\epsilon_{k_2\geq 4})]$ no longer vanish individually, and
relations in the derivation algebra \cite{Tsunogai, Pollack, Broedel:2015hia}
hold non-trivially.

\vspace{-0.15cm}

%%%%%%%%%%%%%%%%%%%%%%%%%%%%%%%%%%%%%%%%%
\section{Examples for initial values}
%%%%%%%%%%%%%%%%%%%%%%%%%%%%%%%%%%%%%%%%%

\vspace{-0.3cm}

\noindent
This section is dedicated to the degeneration of $A$-cycle integrals (\ref{let7}) at the
cusp $\tau \rightarrow i \infty$ which enters the $\alpha'$-expansion \eqref{let6} as
an initial value.

%%%%%%%%%%%%%%%%%%%%%%%%%%%%%%%%%%%%%%%%%%%%%%%%%%%%%
%%%%%%%%%%%%%%%%%%%%%%%%%%%%%%%%%%%%%%%%%%%%%%%%%%%%%
\medskip
{\bf A. Generalities:} The behaviour of $A$-cycle integrals at the cusp is
most conveniently studied in the variables
\beq
\sigma_j = e^{2\pi i z_j} \, , \ \ \ \ \dd z_j = \frac{ \dd \sigma_j }{2\pi i \sigma_j}
\, , \ \ \ \  G_{ij} = i \pi  \frac{\sigma_i {+} \sigma_j}{\sigma_i {-} \sigma_j}\, ,
  \label{let42}
\eeq
where the planar Green function and Kronecker--Eisenstein
series degenerate to ($\sigma_{ji}\equiv \sigma_j{-}\sigma_i$)
\begin{align}
\lim_{\tau \rightarrow i\infty}  \Omega(v_{ij},\eta,\tau) &= \pi \cot(\pi \eta) + G_{ij}   \label{let43} \\
\lim_{\tau \rightarrow i\infty}  {\cal G}(v_{ij},\tau) &= \frac{1}{2} \log (\sigma_i)
+  \frac{1}{2} \log (\sigma_j) -   \log (\sigma_{ji}) \, . \notag
\end{align}
Their non-planar analogues take an even simpler form,
\beq
\lim_{\tau \rightarrow i\infty}  \! \Omega(v_{ij}{+}\tfrac{\tau}{2},\eta,\tau) = \frac{ \pi }{ \sin(\pi \eta) } \, , \ \ \
\lim_{\tau \rightarrow i\infty} \! {\cal G}(v_{ij}{+}\tfrac{\tau}{2},\tau)  = 0\, .
\label{let44}
\eeq
Since string-theory applications of \eqref{let6} involve the coefficients w.r.t.\ $\eta_j$,
we will need the expansions
\begin{align}
 \pi \cot(\pi \eta) &=
 \frac{1}{\eta} - 2 \sum_{k=1}^{\infty} \zeta_{2k} \eta^{2k-1} \label{let45}  \\
 \frac{ \pi }{\sin(\pi \eta)} &=
 \frac{1}{\eta} +  \sum_{k=1}^{\infty} \frac{2^{2k-1} {-}1 }{2^{2k-2}} \zeta_{2k}  \eta^{2k-1}\,. \notag
\end{align}
As will be detailed in \cite{bigpaper}, the $\sigma_j$-integration in $n$-point
$Z^{i \infty}_{\vec{\eta}}$ lines up with explicitly known combinations of
$N=(n{+}2)$-point disk integrals \cite{Zfunctions}
\begin{align}
Z^{\rm tree}&(a_1,a_2,\ldots,a_N|1,2,\ldots ,N) = \! \! \! \!  \! \! \! \! \! \! \! \!\int \limits_{-\infty <\sigma_{a_1} < \sigma_{a_2} < \ldots < \sigma_{a_N}< \infty} \! \! \! \!   \! \! \! \!\notag \\
& \ \   \frac{ \dd \sigma_1 \, \dd \sigma_2\, \ldots \, \dd \sigma_N }{\textrm{vol} \ \textrm{SL}_2(\mathbb R)} \, \frac{ \prod^N_{i<j} |\sigma_{ij}|^{-s_{ij}} }{ \sigma_{12} \sigma_{23} \ldots \sigma_{N-1,N} \sigma_{N,1}}  \, .
 \label{let49}
\end{align}
The two extra punctures $n{+}1\rightarrow +$ and
$n{+}2\rightarrow -$ are associated with Mandelstam invariants
\beq
s_{j+} = s_{j-} = - \frac{1}{2} \sum^n_{1\leq i\neq j} s_{ij} \, , \ \ \ \ \ \ s_{+,-} = \sum_{1\leq i <j }^n s_{ij}\, .
\label{inival3}
\eeq
The $\alpha'$-expansion of (\ref{let49}) and therefore $Z^{i \infty}_{\vec{\eta}}$ involves
multiple zeta values (MZVs) with $n_j \in \NN$,
% NEW EQ
\beq
\zeta_{n_1,n_2,\ldots,n_r} = \! \! \! \! \! \! \! \! \! \sum_{0<k_1<k_2<\ldots <k_r}^\infty \! \! \! \! \! \! \! \! \! k_1^{-n_1}  k_2^{-n_2} 
\ldots  k_r^{-n_r}  \, , \ \ \ \ n_r\geq2 
\, ,
\label{MZV}
\eeq
which can be systematically generated from the all-multiplicity methods
of \cite{Broedel:2013aza, Mafra:2016mcc}. 
% NEW STUFF 
While the four-point expansions are captured by Riemann zeta values at $r=1$,
\begin{align}
&Z^{\rm tree}(1,2,+,- | 1,2,-,+) = - \frac{1}{s_{12}} \label{4ptex}  \\
& \ \ \ \ \times 
\exp\bigg( \sum_{k=2}^{\infty} \frac{ \zeta_k }{k}   \big[ s_{12}^k  + s_{2+}^k - (s_{12}{+}s_{2+})^k \big] \bigg)\, ,
\notag
\end{align}
disk integrals (\ref{let49}) at $N\geq 5$ points additionally involve
MZVs (\ref{MZV}) at higher depth $r\geq2$ \cite{Schlotterer:2012ny}.

%%%%%%%%%%%%%%%%%%%%%%%%%%%%%%%%%%%%%%%%%%%%%%%%%%%%%
%%%%%%%%%%%%%%%%%%%%%%%%%%%%%%%%%%%%%%%%%%%%%%%%%%%%%
\medskip
{\bf B. Two points:} Planar initial values at two points
descend from four-point tree-level integrals,
\begin{align}
Z^{i\infty}_{\eta_2}(1,2|1,2)
&=   \pi \cot(\pi \eta_2)  \,  2i  \sin \Big( \frac{ \pi s_{12}}{2} \Big) \notag \\
&\ \ \ \ \times \int^1_{0} \frac{ \dd \sigma_2}{ 2\pi i \sigma_2}
  \, \sigma_2^{s_{12}/2} (1-\sigma_2)^{-s_{12}}  \label{let46}
\\
&= \pi \cot(\pi \eta_2) \frac{ \Gamma(1-s_{12}) }{ \big[ \Gamma(1-\tfrac{s_{12}}{2}) \big]^2}  \, .
\notag
\end{align}
The factor of $2i  \sin ( \frac{ \pi s_{12}}{2})$ and similar trigonometric functions below stem
from contour deformations detailed in \cite{bigpaper}. The gamma functions with standard $\alpha'$-expansion
\begin{align}
\frac{ \Gamma(1-s_{12}) }{ \big[ \Gamma(1-\tfrac{s_{12}}{2}) \big]^2}&=
\exp \bigg( \sum_{k=2}^{\infty} \frac{ \zeta_k }{k}  ( 1- 2^{1-k}) s_{12}^k  \bigg) \label{let48}  \\
&= 1 {+} \frac{1}{4} s_{12}^2 \zeta_2 {+} \frac{1}{4} s_{12}^3 \zeta_3
{+} \frac{19}{160} s_{12}^4 \zeta_2^2 {+} {\cal O}(\alpha'^5)
 \notag
\end{align}
%
%NEW
arise from the kinematic limit (\ref{inival3}) of (\ref{4ptex}) and
% OLD
do not appear in the non-planar counterpart of (\ref{let46})
\beq
Z^{i\infty}_{\eta_2}\big( \begin{smallmatrix} 2 \\ 1 \end{smallmatrix}|1,2\big) = \frac{ \pi }{\sin(\pi \eta_2)}\, .
\label{let47}
\eeq
%

%%%%%%%%%%%%%%%%%%%%%%%%%%%%%%%%%%%%%%%%%%%%%%%%%%%%%
%%%%%%%%%%%%%%%%%%%%%%%%%%%%%%%%%%%%%%%%%%%%%%%%%%%%%
\medskip
{\bf C. Three points:} Degenerate $A$-cycle integrals at three points introduce
five-point disk integrals,
\begin{align}
&Z^{i\infty}_{\eta_2,\eta_3}(1,a_2,a_3|1,2,3)   \label{let50}  \\
&= \pi^2 \Big(  \cot(\pi \eta_{23})\cot(\pi \eta_3) +\frac{  s_{13}}{s_{123}} \Big) I^{\rm tree}(1,a_2,a_3|1)  \notag \\
& \ \ + \pi \Big(  \cot(\pi \eta_{23})  +  \frac{ s_{23}}{s_{12}}   \cot(\pi \eta_{3}) \Big)
 I^{\rm tree}(1,a_2,a_3|G_{23})\, , \notag
\end{align}
where \small
\begin{align}
 I^{\rm tree}(1,a_2,a_3|1)
 &= - \frac{1}{2\pi^2}
 \Big[ \sin \Big( \frac{ \pi }{2} (s_{1a_2}{+}s_{23}) \Big) \sin \Big( \frac{ \pi }{2} s_{1a_3} \Big)   \notag\\
&\! \! \! \! \! \! \! \! \! \! \! \! \! \! \! \! \! \!  \times
 \Big( Z^{\rm tree}(+,a_2,a_3,1,-| +,2,3,-,1)  \notag \\
 &\! \! \! \! \! \! \! \! \! \! \! \! \! \! \!   + Z^{\rm tree}(+,a_2,a_3,1,-| +,3,2,-,1) \Big) +  (2\leftrightarrow 3)  \Big]
\notag\\
%%%
%%%
%%%
 I^{\rm tree}(1,a_2,a_3|G_{23})&=  \frac{1}{2\pi}    \Big[ \sin \Big( \frac{ \pi }{2} (s_{1a_2}{+}s_{23}) \Big) \cos \Big( \frac{ \pi }{2} s_{1a_3} \Big) \notag \\
 &\! \! \! \! \! \! \! \! \! \! \! \! \! \! \! \! \! \!  \times
 \Big( Z^{\rm tree}(+,a_2,a_3,1,-| +,2,3,-,1)  \label{let51}  \\
 &\! \! \! \! \! \! \! \! \! \! \! \! \! \! \!   - Z^{\rm tree}(+,a_2,a_3,1,-| +,3,2,-,1) \Big) +  (2\leftrightarrow 3)  \Big] \, .\notag
\end{align}
\normalsize
Their leading low-energy orders read \cite{bigpaper}
\begin{align}
 I^{\rm tree}(1,2,3|1)&= \frac{1}{2} + \frac{ \zeta_2 }{8}(s_{12}^2 {+} s_{13}^2 {+} s_{23}^2)  {+} {\cal O}(\alpha'^3) \! \!
\label{let38} \\
 I^{\rm tree}(1,2,3|G_{23})&= \frac{1}{s_{23}}
+  \frac{\zeta_2}{4s_{23}}  (s_{12} {+} s_{13} {+} s_{23})^2 {+} {\cal O}(\alpha'^2)  \notag
\end{align}
and exemplify that integrals over $k$ factors of $G_{ij}$ in (\ref{let42})
may have up to $k$ kinematic poles.

Non-planar three-point initial values in turn boil down to four-point
disk integrals with $\alpha'$-expansions in (\ref{let48}),
\begin{align}
Z^{i\infty}_{\eta_2,\eta_3} \big(\begin{smallmatrix} 3 \\ 1,2 \end{smallmatrix} |1,2,3\big) &
=  \frac{ \pi^2 \cot(\pi \eta_{23}) }{\sin(\pi \eta_3)}  \frac{ \Gamma(1-s_{12}) }{ \big[ \Gamma(1-\tfrac{s_{12}}{2}) \big]^2}
\label{let39} \\
Z^{i\infty}_{\eta_2,\eta_3} \big(\begin{smallmatrix} 3 \\ 1,2 \end{smallmatrix} |1,3,2\big) &
=  \frac{ \pi^2}{ \sin(\pi \eta_{23}) \sin(\pi \eta_2)}  \frac{ \Gamma(1-s_{12}) }{ \big[ \Gamma(1-\tfrac{s_{12}}{2}) \big]^2}
 \,.\notag
\end{align}
%

%hhh
\vspace{-0.15cm}

%%%%%%%%%%%%%%%%%%%%%%%%%%%%%%%%%%%%%%%%%%%%%%%%%%%%%
\section{Conclusions and further directions}
%%%%%%%%%%%%%%%%%%%%%%%%%%%%%%%%%%%%%%%%%%%%%%%%%%%%%%

\vspace{-0.3cm}
\noindent
In this letter we presented a method to expand a generating series of
genus-one integrals \eqref{let7} relevant to one-loop open-string amplitudes.
At each order in the inverse string tension $\alpha'$, our main result (\ref{let6})
pinpoints the accompanying eMZVs in their minimal and canonical representation
via iterated Eisenstein integrals.

Genus-zero integrals relevant to open-string tree amplitudes obey
Knizhnik--Zamolodchikov equations with a characteristic linear factor
of $\alpha'$ on their right-hand side \cite{Broedel:2013aza}. This structure
is analogous to the $\varepsilon$-form of differential equations among Feynman integrals
with dimensional-regularization parameter $\varepsilon$ \cite{Henn:2013pwa,
Adams:2018yfj}, suggesting a correspondence between $\alpha'$ and $\varepsilon$.
By the linearity of the differential operators $D^{\tau}_{\vec{\eta}}$ in $s_{ij}=-2\alpha'k_i\cdot k_j$,
the Knizhnik--Zamolodchikov--Bernard-type equation \eqref{let15} also becomes linear in $\alpha'$.
So our results generalize this intriguing correspondence to genus one
and provide the string-theory analogue of the $\varepsilon$-form for differential
equations of elliptic Feynman integrals \cite{Adams:2018yfj}.

The generating functions $Z^\tau_{\vec{\eta}} $ are expected to comprise any
moduli-space integral in massless one-loop amplitudes of
open bosonic strings and superstrings upon expansion in $\eta_j$. Accordingly,
they are proposed to generalize the universal disk-integrals (\ref{let49}) that
appear in the double-copy representation
of string tree-level amplitudes \cite{Mafra:2011nv, Zfunctions}.
Hence, the study of the genus-one
integrals $Z^\tau_{\vec{\eta}} $ is an essential step towards
universal double-copy structures in one-loop amplitudes of different string theories
that generalize those of the superstring \cite{Mafra:2017ioj}.

The generating functions $Z^\tau_{\vec{\eta}}$ can be adapted to a closed-string
context, encoding the integrals over torus punctures in one-loop amplitudes of
type-II, heterotic and closed bosonic string theories. Closed-string analogues of
$Z^\tau_{\vec{\eta}}$ will be shown \cite{closedstring} to obey similar differential equations
and to shed new light on the properties of modular graph forms \cite{mgforms}
including their relation with open-string amplitudes \cite{openclosed}.

Moreover, the method of this work to infer moduli-space integrals from differential equations
should be applicable at higher loops. In the same way as
disk integrals were used as the initial value for our one-loop results,
higher-genus integrals in string amplitudes
are expected to obey differential equations w.r.t.\ complex-structure moduli
such that their separating and non-separating degenerations set the initial conditions.
It would be interesting to explore a differential-equation approach of this type to the higher-genus
modular graph functions of \cite{highermgf}.

In summary, our new approach to one-loop open-string amplitudes via differential
equations connects with state-of-the-art techniques in particle phenomenology and provides
explicit matrix representations of profound number-theoretic structures. As will be elaborated
in \cite{bigpaper}, our results manifest important formal properties of string amplitudes such as
uniform transcendentality, coaction formulae and the dropout of twisted eMZVs
from non-planar open-string amplitudes.

%%%%%%%%%%%%%%%%%%%%%%%%%%%%%%%%%%%%%%%
%%%%%%%%%%%%%%%%%%%%%%%%%%%%%%%%%%%%%%%
%%%%%%%%%%%%%%%%%%%%%%%%%%%%%%%%%%%%%%%

\bigskip
 \noindent\textit{Acknowledgements:}
We are grateful to Johannes Broedel, Jan Gerken, Axel Kleinschmidt, Nils Matthes and Federico Zerbini
for inspiring discussions and collaboration on related topics. Moreover, Claude Duhr,
Hermann Nicolai, Albert Schwarz and in particular Sebastian Mizera are thanked for valuable discussions,
and we are grateful to Sebastian Mizera for helpful comments on a draft.
We would like to thank the organizers of the programme
``Modular forms, periods and scattering amplitudes'' at the ETH Institute for Theoretical Studies in Z\"urich
for providing a stimulating atmosphere and financial support. 
CRM is supported by a University Research Fellowship from the Royal Society.
OS is grateful to the organizers
of the workshop ``Automorphic Structures in String
Theory'' at the Simons Center in Stony Brook and those of the workshop ``String Theory from a Worldsheet
Perspective'' at the GGI in Florence for setting up inspiring meetings.
OS is supported by the European Research Council under
ERC-STG-804286 UNISCAMP. 

\end{document}